\documentclass[twocolumn,showpacs,pra,floatfix]{revtex4}

\usepackage{graphicx}% Include figure files
\usepackage{dcolumn}% Align table columns on decimal point
\usepackage{bm}% bold math
\usepackage{epstopdf}

\begin{document}

\title{Measurements of trap dynamics of cold OH molecules using resonance enhanced multiphoton ionization}

\author{John M. Gray}
\author{Jason Bossert}
\author{Yomay Shyur}
\author{H. J. Lewandowski}

\affiliation{JILA and Department of Physics, University of Colorado, Boulder, Colorado 80309-0440, USA}

\date{\today}

\begin{abstract}

Trapping cold, chemically important molecules with electromagnetic fields is a useful technique to study small molecules and their interactions. Traps provide long interaction times that are needed to precisely examine these low density molecular samples. However, the trapping fields lead to non-uniform molecular density distributions in these systems. Therefore, it is important to be able to experimentally characterize the spatial density distribution in the trap. Ionizing molecules in different locations in the trap using resonance enhanced multiphoton ionization (REMPI) and detecting the resulting ions can be used to probe the density distribution even with the low density present in these experiments because of the extremely high efficiency of detection. Until recently, one of the most chemically important molecules, OH, did not have a convenient REMPI scheme. Here, we use a newly developed 1 + 1' REMPI scheme to detect trapped cold OH molecules. We use this capability to measure trap dynamics of the central density of the cloud and the density distribution. These types of measurements can be used to optimize loading of molecules into traps, as well as to help characterize the energy distribution, which is critical knowledge for interpreting molecular collision experiments.

\end{abstract}


\maketitle

\section{\label{sec:Introduction}Introduction}

Confining cold molecules in electric, magnetic, or optical traps allows for precision studies of molecules and their interactions due to long interrogation times. Neutral molecule trap lifetimes can range from seconds to minutes \cite{RempeTrap,DecTrapOH2005,Doylelongtrap}, depending on the residual background gas pressure and blackbody optical pumping rates \cite{Blackbodypumping}. These long interaction times are required to study molecular interactions in cold systems, as the density of trapped samples is typically very low ($10^{3}-10^8$ cm$^{-3}$)  \citep{DecTrapOH2005,RempeTrap,FirstTrap2000,DecelTrapND32002}. The long lifetimes increase the probability that interactions can take place before the molecules are ejected from the trap by collisions with background gas or by photon scattering \cite{Blackbodypumping}. There have been many recent experiments that have explored various electrostatic  \cite{DoubleWellTrap,SplitCapTrap,CH3Ftrap,Rieger2005} and magnetic \cite{DoyleMag1,MethyRadMag,Shtrapping} trapping geometries, and others that demonstrated additional in-trap cooling \cite{TWSDtrap1,TWSDtrap2,TWSDtrap3,Zeppenfeld2012,Glockner2015,Prehn2016}. Importantly, several experiments have taken advantage of trapped molecules to study the properties of the molecules themselves, including vibrational relaxation of OH \cite{OHviblifetime} and NH \cite{NHviblifetime}. Additonally, studies of molecular interactions have benifited from trapped samples including cold atom-molecule collisions \cite{Parazzoli, DoyleColl, DoyleAtomMol}, molecule-molecule collisions \cite{OHND3collisions}, and collisionally assisted evaporative cooling \cite{OHOHcollisions}. (Here, we restrict our discussions to cold molecules not created by assembly of ultracold atoms.)

One of the most challenging aspects of these experiments is detecting the trapped molecules. The two main methods are laser induced florescence (LIF), where emitted photons are detected using a photo-multipler tube, and resonant ionization, where the resulting ions are detected using a microchannel plate (MCP) detector. Resonant ionization typically has a higher efficiency, as ions can be detected with nearly unit efficiency, while photon collection in LIF is limited by the solid angle of the detection optics, which is only a few percent at best.  LIF detection not only produces a lower signal, but can also have a larger noise contribution, as the molecules are trapped close to reflective surfaces, which increases the scattered light impinging on the detector. For ion detection, the ions are usually sent through a time-of-flight mass spectrometer (TOFMS), nearly eliminating the contribution to the noise from other molecular species. Finally, because of the low signal-to-noise (S/N) ratio, experiments that use LIF must illuminate the entire cloud of molecules to increase the number of photons emitted by the molecules. These measurements produce information about only the total trapped number or average density and not about the spatial distribution of the moleules in the trap. Resonance enhanced multiphoton ionization (REMPI), on the other hand, has a high enough S/N ratio that the ionization laser can be focused to illuminate a small portion of the trapped sample. Then, by making measurements of the molecular density at different laser positions, we can profile the density distribution in the trap. The density distribution in the trap can be used to determine the energy distribution or temperature, which are important parameters to measure for many cold molecule experiments. The importance of measuring the density distribution, and not just the number in the trap, was shown in the study of  collisions between electrostatically trapped ammonia molecules and magnetically trapped rubidium atoms \cite{Parazzoli}. Since the density, and thus the collision rate, varies spatially in the trap, extracting collision cross-sections requires a detailed understanding of the molecular distribution. Therefore, studies of trapped molecules benefit greatly from an efficient and spatially selective resonant ionization scheme. 

Convenient REMPI schemes have been used to measure the density distribution and understand collisional dynamics with trapped ammonia \cite{Parazzoli}. However, until recently, the OH radical, which is a fundamental molecule in atmospheric \cite{AtmosphericOH}, interstellar medium \cite{InterstellarGuibert,InterstellarYusef}, and combustion chemistry \cite{CombustionSmith,CombustionVora}, did not have an efficient REMPI scheme and all trapped studies used LIF \citep{DecTrapOH2005,OHND3collisions,OHOHcollisions,OHviblifetime,JunOHtrapping,JunOHtrappedHeD2}. In 2011, the Lester group demonstrated a new UV + VUV REMPI scheme to state-selectively ionize ground-state OH radicals using 118 nm light as the VUV photon \cite{OHREMPI2011,OHREMPI2014,OHREMPI2016}. They used the REMPI scheme to examine the OH products from unimolecular dissociation of Criegee intermediates \cite{Criegee1, Criegee2}. Here, we present the first measurements of trapped OH molecules using this ionization detection scheme. This method allows us to measure not only the relative number of molecules in the trap, but also the temporal evolution of the spatial density distribution. We use this new capability to measure trap loading dynamics to optimize the loading process and determine the energy distribution in the tap. Proper trap loading and characterization of the energy distribution are important for future atom-molecule collision experiments.

\section{\label{sec:REMPI} REMPI of OH}

\begin{figure}
\includegraphics[width=3.25in]{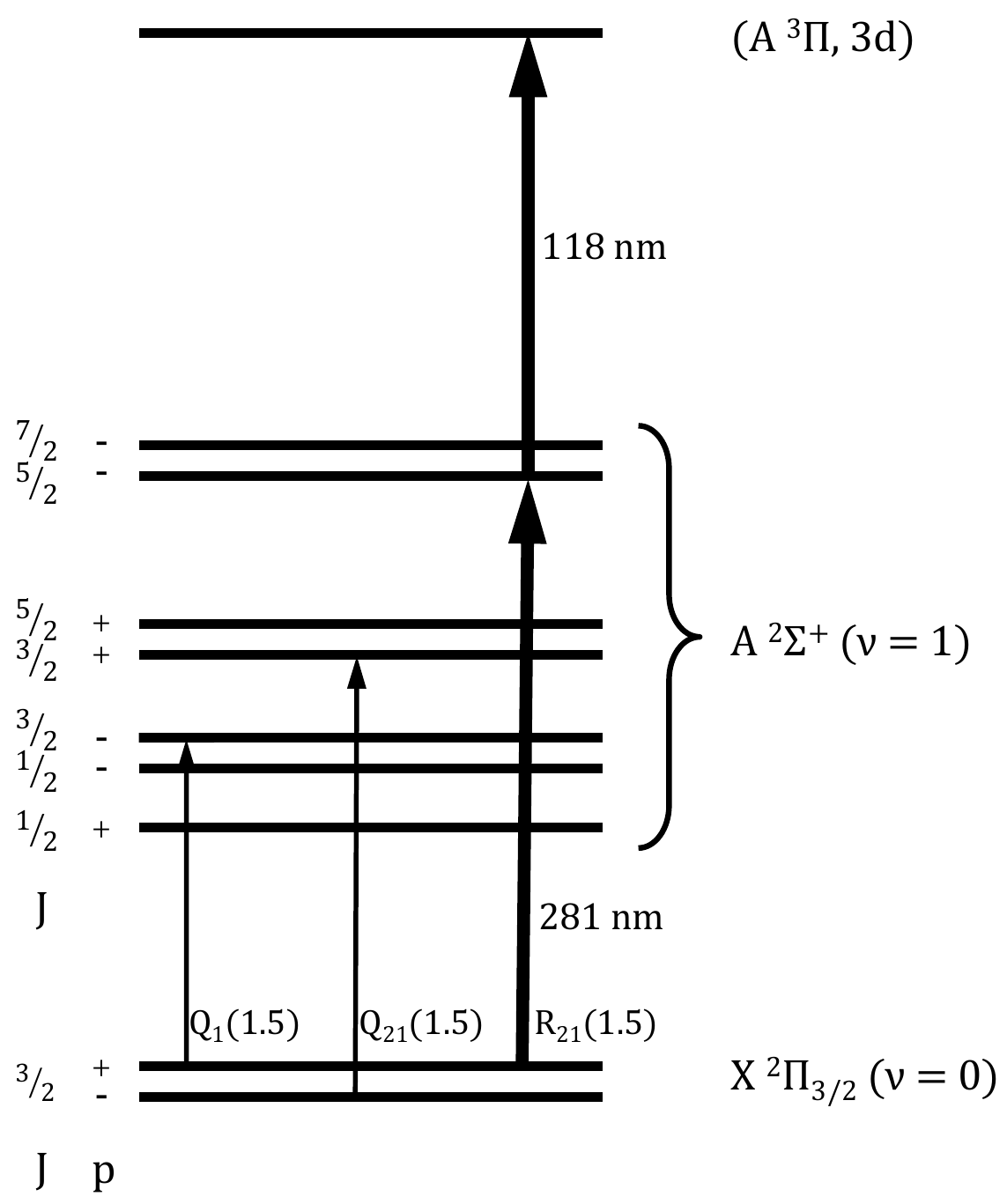}%
\caption{\label{fig:EnergyLevels}Energy level diagram of OH showing the relevant levels for the 1 + 1' REMPI detection scheme. The lowest two levels are the $\Lambda$-doublet states of $X^2\Pi_{3/2}$. The first photon is resonant with the $A^2\Sigma^+, \nu = 1 \leftarrow X^2\Pi_{3/2}, \nu = 0$ transition and is denoted by R$_{21}$(1.5). The second photon transfers the molecule into the $A^3 \Pi, 3d$ Rydberg state, from which it autoionizes. $J$ is the total angular momentum, and $p$ is the parity. Vertical spacing between energy levels is not to scale.}
\end{figure}

The 1+1' REMPI scheme used to detect trapped OH molecules uses two photons; the first photon causes a bound-to-bound transition around 281 nm and the second photon (118 nm) excites the molecule to an autoionizing Rydberg state whose lifetime is on the order of a picosecond \citep{OHREMPI2016}. The relevant energy structure of OH is shown in Fig. \ref{fig:EnergyLevels}. The trapped molecules are in the ground rotational ($J = 3/2$)  and vibrational ($\nu = 0$) state of the $X^2\Pi_{3/2}$ electronic state. The first photon (UV) is tuned to the $A^2\Sigma^+, \nu = 1 \leftarrow X^2\Pi_{3/2}, \nu = 0$ transition. There are several possible rotatonal states to tranfer the molecule into with the UV photon. The stongest A $\leftarrow$ X transision is Q$_1$ (1.5) shown in Fig. \ref{fig:EnergyLevels}. Beames et. al  observed that in a supersonically expanded jet of OH, the highest ionization signal was achieved when the UV photon was resonant with the Q$_{21}$(1.5) transition \cite{OHREMPI2014}. However, this transition originates from the lower $\Lambda$-doublet level and only the upper $\Lambda$-doublet level is trappable with the static electric fields used in our experiments. The strongest transition that originates from upper $\Lambda$-doublet level is the R$_{21}$(1.5) transition, which is used in the work presented here.  The second step in the process transfers the molecule to an autoionizing Rydberg state using a 118 nm photon.

\section{\label{sec:Exp} Experimental Setup}

\begin{figure}
\includegraphics[width=3.25in]{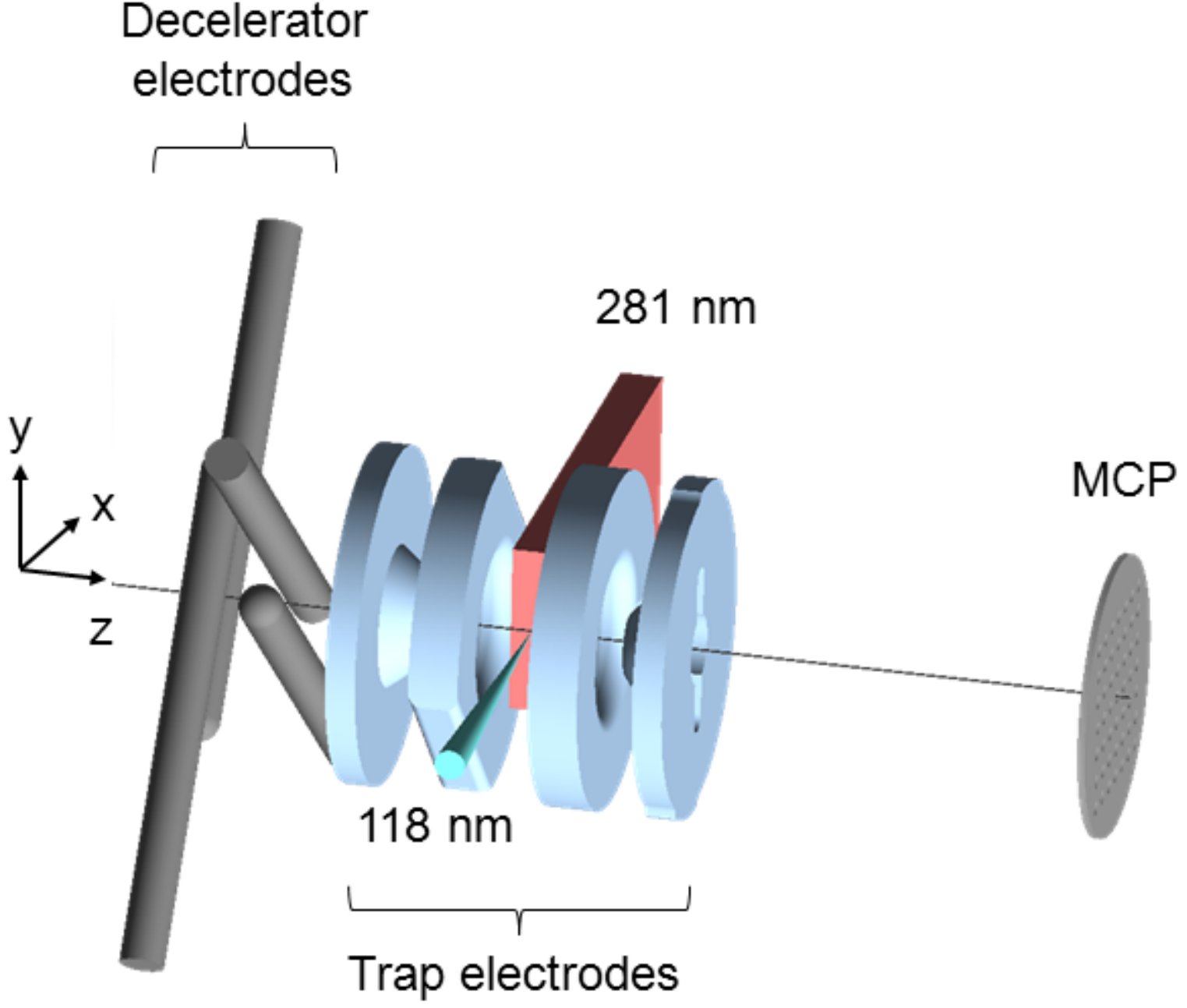}%
\caption{\label{fig:ExpSetup} Experimental setup showing the last two rod pairs of the Stark decelerator, four trap electrodes, detection lasers, and MCP detector. The red retangular prizm represents the 281 nm laser beam, while the blue cone represents the 118 nm light. The molecular beam propagates along the z-axis. (Not to scale)}
\end{figure}

\begin{figure}
\includegraphics[width=3.25in]{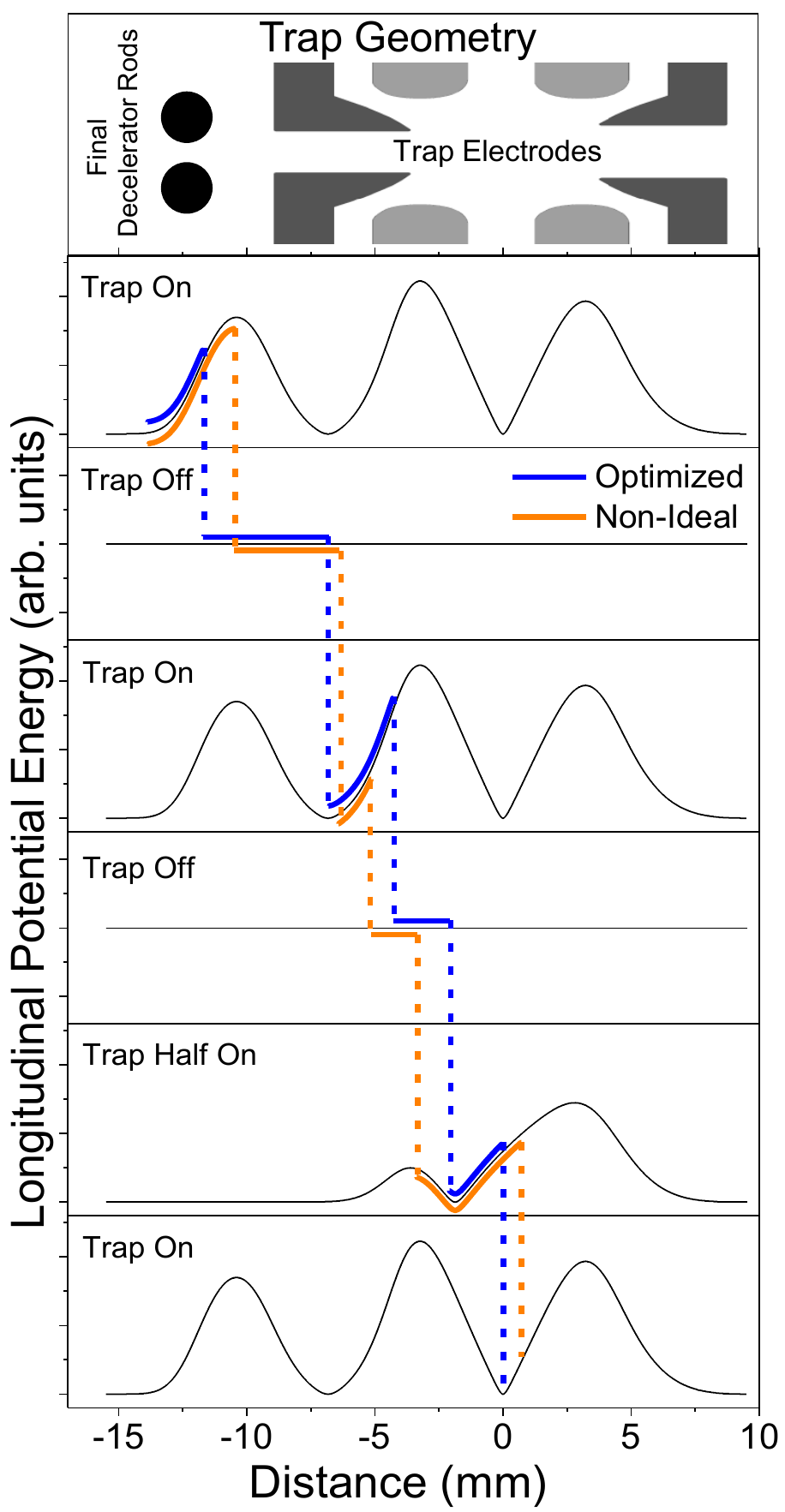}%
\caption{\label{fig:TrapSequence}Diagram of trap loading process for both an optimized and a non-ideal sequence. The location of the final decelerator electrode and the four trapping electrodes are shown at the top. The longitudinal potential energy curves seen by the OH molecules are shown in the panels for three different configurations of voltage applied. The ``Trap on'' configuration has the following voltages applied in order from left to right (decelerator, left trap endcap, left trap ring, right trap ring, right trap endcap) = (0, -10, +10, +10, -10) kV. ``Trap Half on'' has the following voltage configuration (0, 0, +10, 0, -10) kV, and ``Trap off'' has all electrodes grounded. The blue (top) curve represents the potentials seen by the central molecule in the optimized sequence and the orange (bottom) curve represents the non-ideal sequence. The switches between configurations are shown by vertical lines.}
\end{figure}

 The experimental setup is shown in Fig. \ref{fig:ExpSetup}. OH radicals are produced at the exit of a piezoelectric transducer (PZT) valve by electric discharge of ~1\% water seeded in krypton.  The molecules are cooled both internally and translationally in the supersonic expansion, producing a beam of OH radicals moving with a mean velocity of 415 m/s and mostly in their rovibronic ground state. The molecular beam is slowed to 34 m/s by a pulsed Stark decelerator, which uses time varying inhomogeneous electric fields to decelerate polar molecules \cite{FirstDeceleration}. An electrostatic trap located at the end of the decelerator is used to reduce the mean velocity of the molecules to nearly zero by switching the potentials on the four trap electrodes on and off at different times during the loading process, (Fig. \ref{fig:TrapSequence}). The final potentials on the trap electrodes form a 3D trap for OH with a trap depth of 700 $\mu$K in the z direction and 250 $\mu$K in the x and y directions (Fig. \ref{fig:ExpSetup}).
 
Molecules in the trap are detected using the 1+1' REMPI scheme. The two detection laser beams enter the center of the trap perpendicular to the molecular beam axis through a 3 mm gap between the two central trap electrodes. A frequency-doubled pulsed dye laser, which produces light at 281 nm, excites the OH molecules on the R$_{21}$(1.5) transition as shown in Fig. \ref{fig:EnergyLevels}. The 281 nm beam has a beam waist of 2 mm in the y direction and is focused by a 50 mm focal length cylindrical lens to have a beam waist of around 0.5 mm in the z direction at the trap center. The pulse energy of this laser is 4.7 mJ per pulse, which is enough to saturate the transition. 

To create the 118 nm light, the third harmonic of a pulsed Nd:YAG laser at 355 nm is frequency tripled in a gas cell containing xenon and argon \cite{118Lockyer1997,Harris118,Our118paper}. The 355 nm light is focused with a 500 mm focal length lens into the gas cell containing 24 torr of xenon mixed with argon at an Ar:Xe ratio of 10.3:1. The xenon provides the non-linear medium, while the argon is used to satisfy the phase-matching condition. We estimate that we create 1 nJ/pulse of 118 nm light. A 250 mm focal length MgF$_2$ lens, which serves as the vacuum window and separates the gas cell from the ultrahigh vacuum trapping chamber, focuses the 118 nm light to a diameter estimated to be 30 $\mu$m at the center of the trap. The chromatic dispersion of the MgF$_2$ lens is used to collimate the residual 355 nm light, the majority of which is blocked by an alumina aperture before the trap. The position of the focus of the 118 nm laser can be scanned both vertically and horizontally using a computer-controlled closed-loop picomotor (New Focus Model 8310). 

The central molecular column density and density distribution along the y-axis is measured during the trap loading process, after the trap has been loaded, and after the trap has been turned off to allow the cloud of molecules to expand. During ionization, the potentials on the four trap electrodes are rapidly switched off during the laser pulses, and then changed to (3, 3, 2, 0) kV to create a uniform electric field and accelerate the ions through a TOFMS to an MCP detector. The signal from the MCP corresponding to mass/charge = 17 is recorded via a multichannel scaler (Stanford Research Systems SR430). For all experimental data presented below, each point represents 1000 individual laser shots (100 seconds of integration) with error bars representing the statistical uncertainty in the measurement.

\section{\label{sec:Exp} Molecular Dynamics Simulations}
To help interpret the experimental results, three-dimensional molecular dynamics simulations are performed in two stages. The first stage simulates the trajectory of the molecules from the nozzle of the PZT valve to the end of the Stark decelerator. The molecules experience time-varying forces due to the interaction of the molecules' electric dipole moment and the time-varying electric field created by the high-voltage electrodes.  More information on this simulation can be found in Refs. \cite{Parazzoli,CollisionsPaper}. Molecules are then propagated through the trapping region using a second simulation using the phase-space distribution of the molecules from the decelerator simulation as the input. The vector force fields for each configuration of the potentials on the electrodes are generated via a commercial finite element solver, COMSOL, and linearly interpolated to the location of each molecule at every time step. A second-order, symplectic, velocity-Verlet integrator is used to propagate the molecules forward in time. A integration time-step of 2 $\mu$s is used to minimize computation time without sacrificing required accuracy. To mimic experimental measurement conditions, every 50$\mu$s, we create a 2D histogram of the number of molecules in the y-z plane, where each bin of the histogram is 50 $\mu$m wide in the y- and z-dimensions and is 6 mm long in the x-dimension. This simulated measurement scheme allows us to record data with high spatial and temporal resolution to compare with experimental data. All trapping simulations were run with an initial number of 3.5$\times 10^6$ molecules.

\section{\label{sec:Results} Results and Discussion}

\subsection{Trap loading dynamics} 

Efficiently loading cold molecules from a decelerator into a trap is critical to be able to perform subsequent experiments, as the number density exiting a decelerator is already low. We use information about the time evolution of the central molecule column density and the in-trap density distribution to optimize loading into a quadrupole electrostatic trap. Previously, only the total number was used in this optimization procedure, which could cause situations where the central column density of the sample was not at a maximum, leading to poor initial conditions for collision studies. 

\begin{figure}
\includegraphics[width=3.25in]{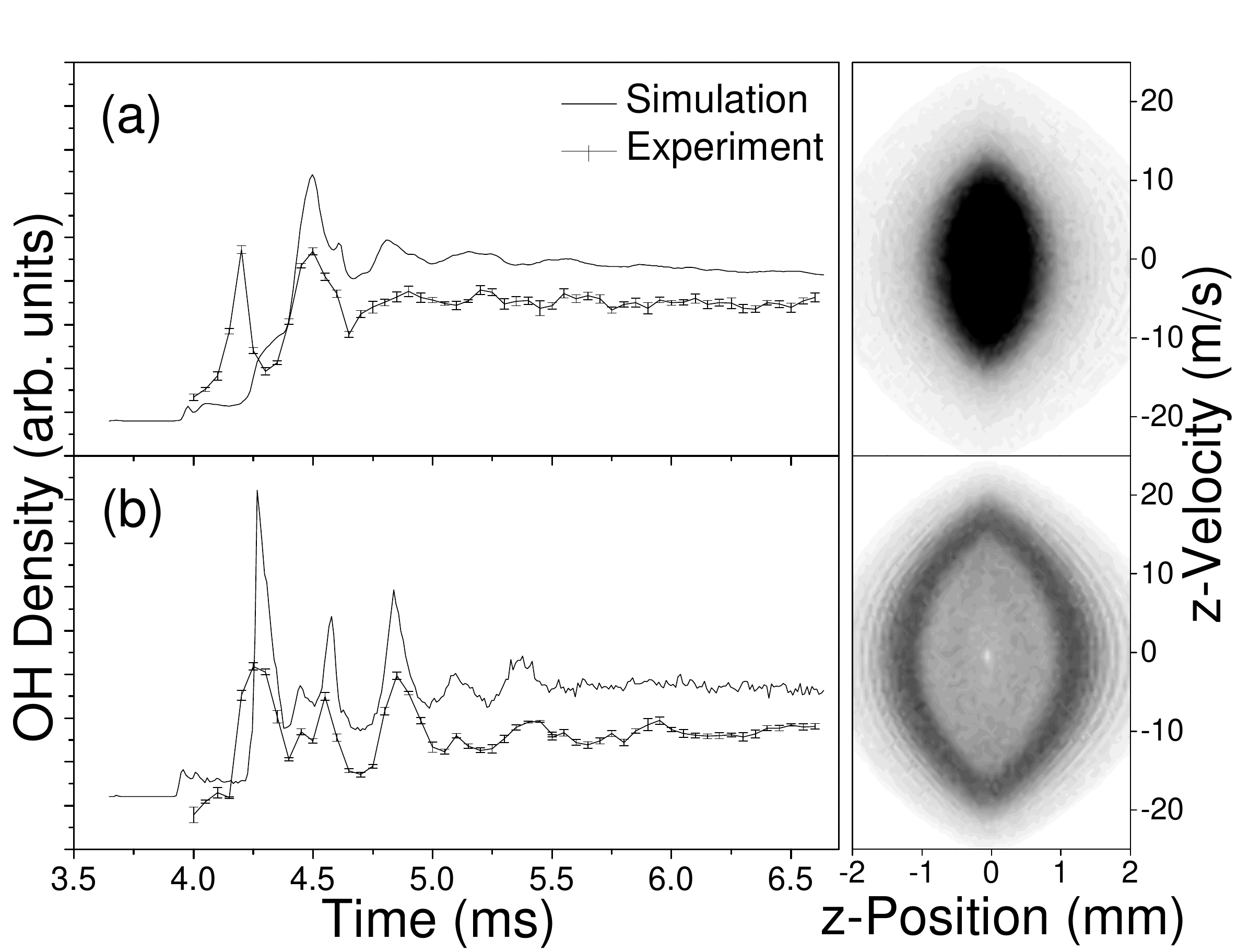}%
\caption{\label{fig:SigvTime} Measured central column density during the trap loading process with corresponding molecular dynamics simulations. The measured central density is shown with error bars connected by lines to guide the eye and the simulations results are shown with solid lines. The simulation results are artificially displaced upward for clarity. (a) Results from the optimized loading scheme. The final trap configuration is turned on at 4.34 ms. (b) Results from the non-ideal loading scheme. The final trap configuration is turned on at 4.30 ms.  The plots at the right correspond to the simulated longitudinal phase-space distribution after 15 ms of trapping.}
\end{figure}

Figure \ref{fig:SigvTime} shows the column density, which was produced by integrating through the x-dimension, at the center of the trap during the loading process for two sample loading procedures (See Fig. \ref{fig:TrapSequence}). Fig. \ref{fig:SigvTime}(a) shows the optimized loading process, while Fig. \ref{fig:SigvTime}(b) shows a process that, although it loads roughly the same number of molecules into the trap, it loads the molecules into a state with higher average energy and lower density at the center of the trap. The results of molecular dynamics simulations are also shown for these loading schemes, and show good agreement of the time evolution of the central column density. Small mismatches between the results of the simulations and the experimental measurements are likely due to the small inaccuracies in the initial molecular distribution used for the trapping simulations. The simulated final phase-space distribution in the trap for the two loading schemes is shown at the right of Fig. \ref{fig:SigvTime}. For the optimized loading scheme, the most dense portion of the phase-space distribution resides at low energies, whereas the molecules are at higher energies in the non-ideal loading scheme. The non-ideal loading scheme produces a distribution that would be unsuitable for collision experiments due to its lower number of molecules in the center of the trap and its higher most probable velocity.

\begin{figure}
\includegraphics[width=3.25in]{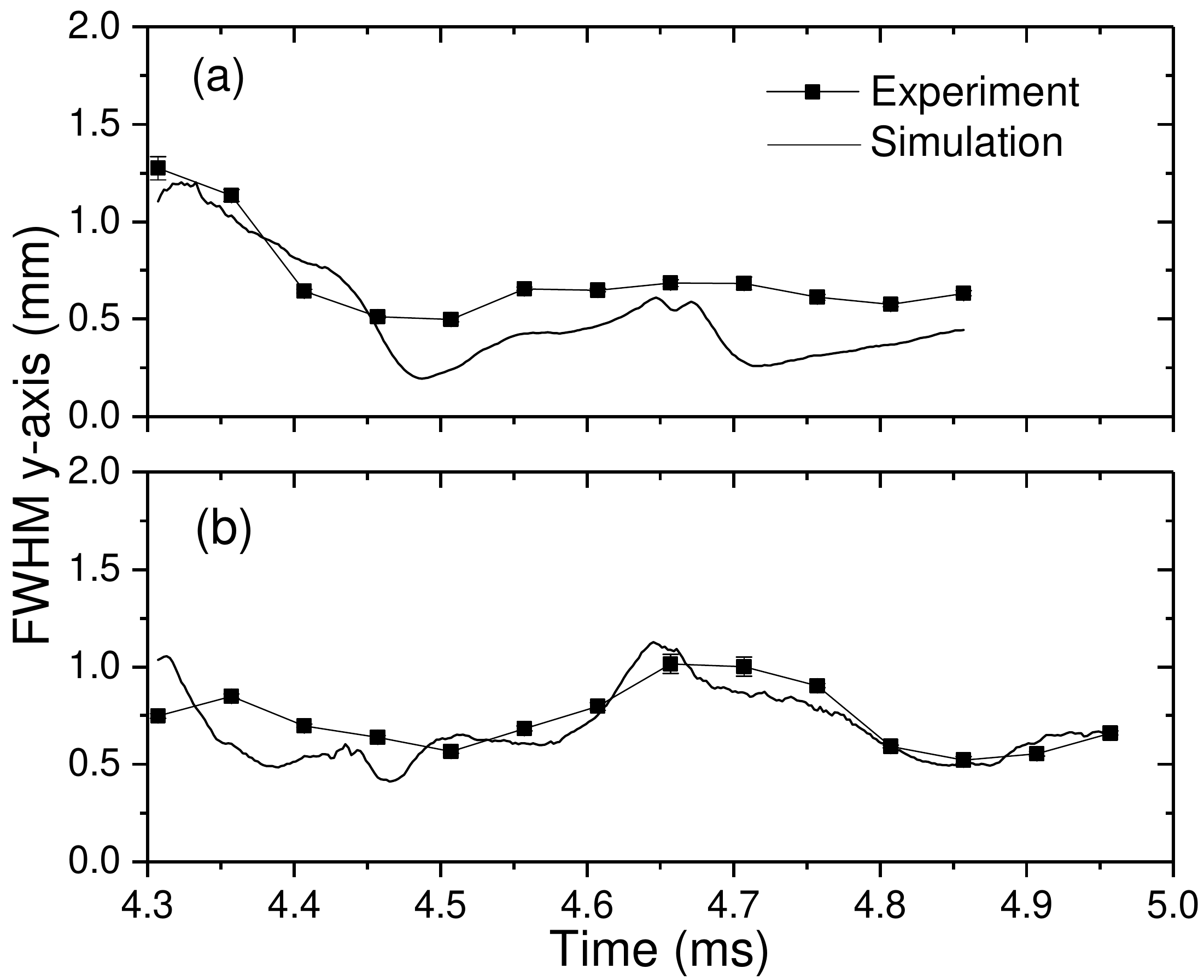}%
\caption{\label{fig:Widths}Width of the trapped OH cloud as a function of time for the (a) optimized loading and (b) non-ideal loading scheme.  The squares represent the experimental measurements, while the solid lines represent simulation results. For both cases, the simulations reproduce the modulation in the width of the cloud after it is trapped. The optimized loading scheme results in fewer oscillations in the cloud width. The statistical uncertainty is on the order of the point size or smaller.}
\end{figure}

\begin{figure}
\includegraphics[width=3.25in]{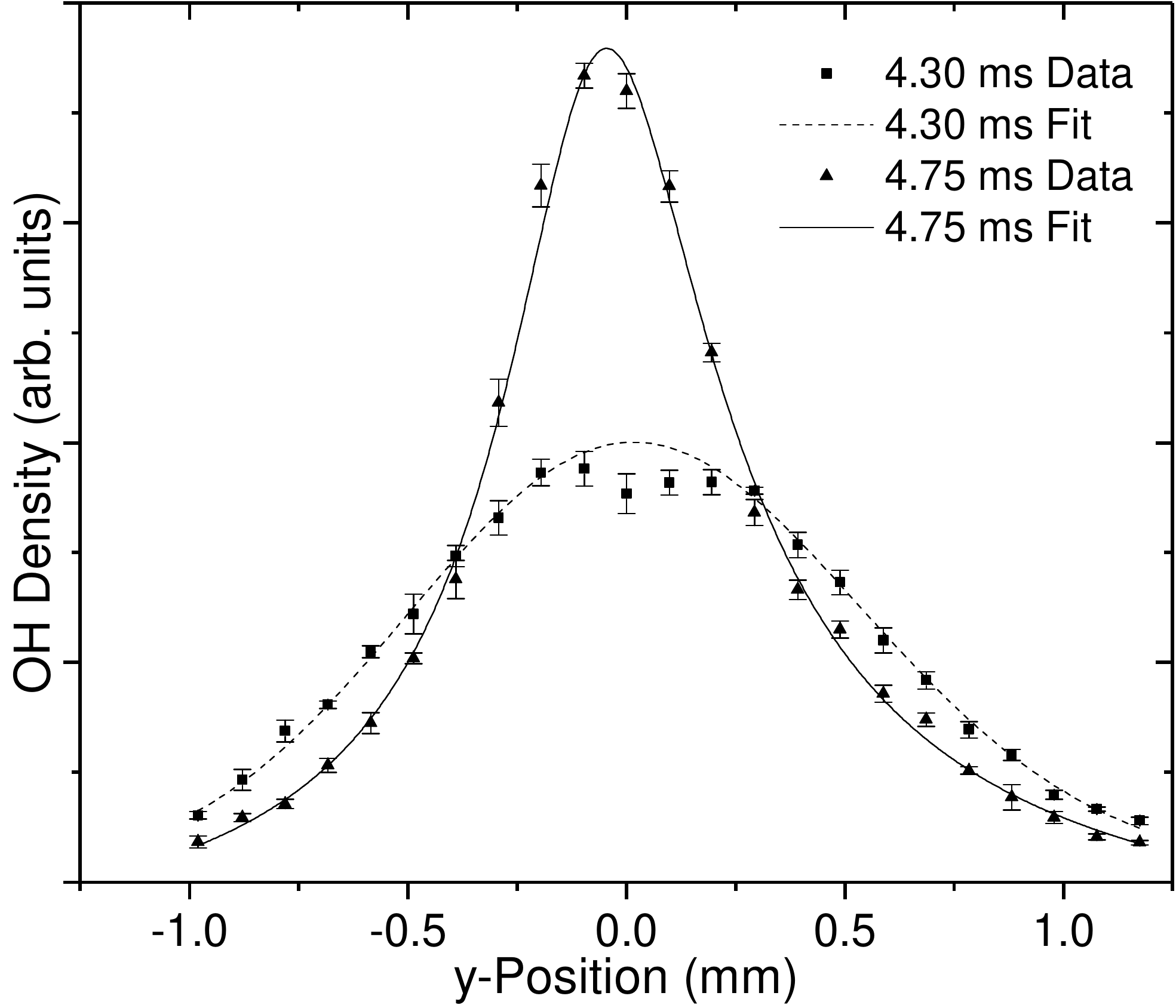}%
\caption{\label{fig:DensityProfile} Measured density profiles of the OH molecules in the trap using the non-ideal loading sequence. These data were recorded directly after the trap was loaded (4.30 ms) and at 450 $\mu$s later (4.75 ms). The lines are Pearson IV distribution fits to the experimental data. The widths of the experimental density profiles are determined by extracting the FWHM of a Pearson IV distribution fit. }
\end{figure}

	In addition to measuring the peak density as a function of time for the loading, we can also measure the dynamics of the density distribution along the y-axis as shown in Fig. \ref{fig:Widths}. Here, the widths are determined by measuring the ions created as the 118 nm laser is scanned along the y-axis of the trap. Since the trapping potentials are not well approximated by either a harmonic or linear potential, an analytic model for the distribution can not be derived from our trapping potentials, so a Pearson IV function was chosen to describe the distribution \cite{Jeffreys1998}. The widths of the experimental and simulated distributions were found by fitting to a Pearson IV function and calculating the full width at half maximum (FWHM) from the fit parameters (Fig. \ref{fig:DensityProfile}). 
	
	For the optimized loading scheme, there are few oscillations in the transverse cloud width or the peak density. However, even after a ms in the non-ideal case, there are observable oscillations in both the cloud width and peak density, indicating the cloud was loaded with excess energy.

\subsection{Ballistic Expansion of OH cloud} 

\begin{figure}
\includegraphics[width=3.25in]{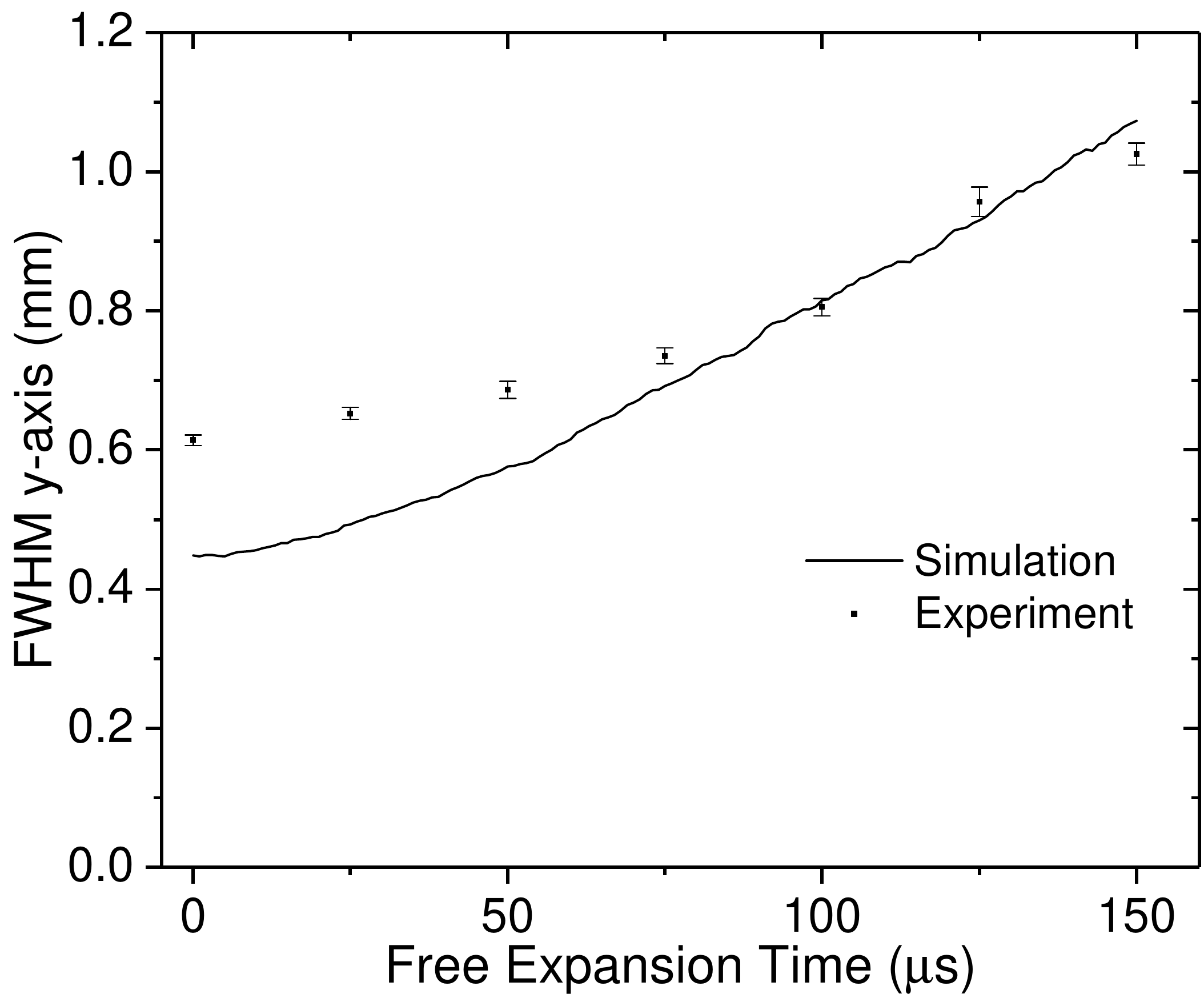}%
\caption{\label{fig:Expansion} Width of the cloud as it expands after the trap is turned off. The molecules were trapped using the optimized loading scheme and held for 15 ms before release in both the experiment and simulation. The initial size of the cloud in the trap differs between the experiment and simulations. This is likely due to the lack of knowledge of the initial distribution of molecules leaving the decelerator. }
\end{figure}

With the ability to measure the density profile, the energy distribution can be  determined experimentally from the expansion of the molecules after the trapping potentials have been turned off. Although this method of determining the temperature is standard practice in cold atom experiments, it has never been used in cold molecule experiments. Figure \ref{fig:Expansion} shows the change in the width of the cloud after the trap turns off for the optimized loading scheme after 15 ms of trapping. We fit the width of the cloud as a function of time to  the following model $\sigma_{y}(t)=\sqrt{(\sigma_{y}(t=0))^2 + (V_{rms} t)^2}$, where $\sigma_{y}(t=0)$ and $V_{rms}$ (root-mean-square velocity) are fitting parameters, and t is the time after the trap is turned off. Fitting $\sigma_{y}(t)$ to the experimental and simulated results gives transverse $V_{rms}$  of 5.5 $\pm$ 0.2 m/s and 6.2 $\pm$ 0.2 m/s respectively. This model assumes the cloud is in thermal equilibrium and is contained in a harmonic trapping potential. Although neither of these assumptions are satisfied by our system, we use this model to determine a reasonable value for the $V_{rms}$. Additionally, one can calculate the rms velocity using the initial size in the trap and the linear potential energy gradient, while employing the virial theorem. Using this, the rms velocity for the  experimental measurements (5.92 $\pm$ 0.07 m/s) and results of simulations (5.13 $\pm$ 0.02 m/s). These measured velocities correspond to a temperature of 50 mK. All the uncertainties stated are statistical. Because the system does not satisfy the conditions of thermal equilibrium or harmonic trapping, the systematic uncertainties likely dominate over the statistical ones. However, all measurements are self-consistent, leading to reasonable confidence in the results.

\section{\label{sec:Conclusion} Conclusion}

Cold, trapped samples of OH molecules could be a useful starting point to explore molecular interactions and cold chemistry. Being able to characterize the trapped samples beyond just total trapped number will be important for these studies.   Using a recently developed 1 + 1' REMPI scheme, we have been able to measure the non-uniform density distribution in a trapped sample of OH. This detailed information has allowed us to optimize the loading of OH into an electrostatic trap. Additionally, the new measurement technique provides information about the trapped molecules' energy distribution from the in-trap cloud size and from the rate of expansion after the trapping fields have been turned off. Previously, these dynamics of trapped OH molecules have been inferred from simulations, but never before observed in experiments.
 
Future work utilizing this detection method includes studying collisions between OH and ultracold Rb atoms. Measuring the density distribution of trapped OH molecules will allow for experimental determination of elastic and inelastic scattering cross-sections in the cold molecule-ultracold atom system. These measured cross-sections will help to improve theoretical models of atom-molecule collisions in the 100 mK temperature regime.

\begin{acknowledgments}
We thank T. Briles for initial development of the system to create 118 nm light and S. Y. T. van de Meerakker for useful discussions about loading electrostatic traps. This work was funded by NSF Grant No. PHY-1125844.
\end{acknowledgments}

\bibliography{TrapOH118BIB}

\end{document}